\documentstyle[preprint,aps]{revtex}
\draft

\begin{document}
\title{Convexity and translational invariance constraint on the
exchange-correlation functional.}
\author{Daniel Joubert}
\address{Physics Department, University of the Witwatersrand, PO Wits 
2050,\\
Johannesburg, South Africa}
\author{Mel Levy}
\address{Department of Chemistry and Quantum Theory Group, Tulane 
University, New\\
Orleans, Louisiana 70118}
\maketitle

\begin{abstract}
Knowledge of the properties of the exchange-correlation functional in 
the
form $\frac 1\lambda v_{xc}([\rho _\lambda ],\frac{{\bf r}}\lambda 
)$, where 
$\rho _\lambda ({\bf r})=$ $\lambda ^3\rho (\lambda {\bf r}),$ is 
important
when expressing the exchange-correlation energy as a line integral $%
E_{xc}[\rho ]=\int_0^1d\lambda \int d{\bf r}\frac 1\lambda 
v_{xc}([\rho
_\lambda ],\frac{{\bf r}}\lambda )\left[ 3\rho ({\bf r})+{\bf 
r.\nabla }\rho
({\bf r})\right] $ (van Leeuwen and Baerends, Phys. Rev. A {\bf 51}, 
170
(1995)). With this in mind, it is shown that in the low density limit 
$%
\lim_{\lambda \rightarrow 0}\int \rho ({\bf r})\nabla ^2\frac 
1\lambda
v_{xc}([\rho _\lambda ],\frac{{\bf r}}\lambda )\ d^3r\leq 4\pi \int 
\rho (%
{\bf r})^2d^3r.$ This inequality is violated in the local-density
approximation.
\end{abstract}


According to density functional theory the electronic groundstate 
energy $E_0
$ can be expressed as 
\begin{equation}
E_0=T_s[\rho ]+U[\rho ]+E_{xc}[\rho ]+\int d{\bf r\ }\rho ({\bf 
r})v({\bf r})
\label{eq1}
\end{equation}
where $v({\bf r})$ is the external potential, $T_s[\rho ]$ is the
non-interacting kinetic energy, $U[\rho ]$ is the classical
electron-electron repulsion energy, and $E_{xc}[\rho ]$ is the
exchange-correlation energy and $\rho ({\bf r})$ is the 
charge-density
distribution \cite{hks,wp,dg}. In a recent paper van Leeuwen and 
Baerends 
\cite{vanLeeuwen} discussed energy expressions in density-functional 
theory
using line integrals. The exchange-correlation energy, for example, 
can be
expressed as 
\begin{equation}
E_{xc}[\rho _2]-E_{xc}[\rho _1]=\int_0^1d\lambda \int d{\bf 
r}\frac{\delta
E_{xc}[\gamma (\lambda )]}{\delta \rho ({\bf r})}\frac{d\gamma 
(\lambda )}{%
d\lambda }{\bf \ }  \label{eq2}
\end{equation}
where $\gamma (0)=\rho _1({\bf r})$ and $\gamma (1)=\rho _2({\bf 
r}).$ For
practical reasons, it is appealing to choose \cite{vanLeeuwen} 
\begin{equation}
\gamma (\lambda )=\rho _\lambda ({\bf r})=\lambda ^3\rho (\lambda 
{\bf r})
\label{eq3}
\end{equation}
Levy \cite{levylow} has shown that in the low density limit, $\lambda
\rightarrow 0,$%
\begin{equation}
E_c[\rho _\lambda ]=\lambda D[\rho ]+...  \label{eq4}
\end{equation}
where $D[\rho ]$ is finite and independent of $\lambda .$ Further, 
since $%
E_x[\rho _\lambda ]=\lambda E_x[\rho ]$\cite{levyperdew85} it follows 
that 
\[
\lim_{\lambda \rightarrow 0}E_{xc}[\rho _\lambda ]=0.
\]
With $v_{xc}([\rho ],{\bf r})=\frac{\delta E_{xc}[\rho ]}{\gamma \rho 
({\bf r%
})},$ the exchange-correlation energy can thus be written as 
\begin{eqnarray}
E_{xc}[\rho ] &=&\int_0^1d\lambda \int d^3r\ v_{xc}([\rho _\lambda 
],{\bf r}%
)\left[ 3\lambda ^2\rho (\lambda {\bf r})+\lambda ^3{\bf r.\nabla 
}_{\lambda 
{\bf r}}\rho (\lambda {\bf r})\right]   \\
&=&\int_0^1d\lambda \int d^3r\ \frac 1\lambda v_{xc}([\rho _\lambda 
],\frac{%
{\bf r}}\lambda )\left[ 3\rho ({\bf r})+{\bf r.\nabla }\rho (\lambda 
{\bf r}%
)\right] .  \label{eq5}
\end{eqnarray}
Also see discussion by van Leeuwen and Baerends below Eq. (20) in 
Ref. \cite
{vanLeeuwen}.

It is therefore of considerable interest to have information on the 
behavior
of $\frac 1\lambda v_{xc}([\rho _\lambda ],\frac{{\bf r}}\lambda )$ 
as a
function of $\lambda .$ In a recent paper Levy and Perdew 
\cite{levyperdew93}
showed that the functional 
\begin{equation}
A[\rho ]=\lim_{\lambda \rightarrow 0}\lambda ^{-1}E_{xc}[\rho 
_\lambda
]+U[\rho ]  \label{eq6}
\end{equation}
is convex, in other words 
\begin{equation}
\int f({\bf r})\frac{\delta ^2A[\rho ]}{\delta \rho ({\bf r})\delta 
\rho (%
{\bf r}^{\prime })}f({\bf r}^{\prime })\ d^3rd^3r^{\prime }\geq 0
\label{eq7}
\end{equation}
for arbitrary $f({\bf r})$ that satisfies 
\begin{equation}
\int f({\bf r})d^3r=0.  \label{eq8}
\end{equation}
In particular, for a finite system or for a system where the charge 
density
vanishes at the boundary, $f({\bf r})=\frac{d\rho ({\bf r})}{dx_i},$ 
($%
x_i\equiv x,y,z)$ satisfies Eq. (\ref{eq8}). Hence 
\begin{equation}
\int \frac{d\rho ({\bf r})}{dx_i}\frac{\delta ^2A[\rho ]}{\delta \rho 
({\bf r%
})\delta \rho ({\bf r}^{\prime })}\frac{d\rho ({\bf r}^{\prime })}{%
dx_i^{^{\prime }}}\ d^3rd^3r^{\prime }\geq 0.  \label{eq9}
\end{equation}

For any functional $Q[\rho ({\bf r})]$ invariant under translation, 
it
follows that $Q[\rho ({\bf r})]=Q[\rho ({\bf r+}t{\bf R})]$ where 
${\bf R}$
is an arbitrary vector and $t$ a scalar. As a consequence \cite{dpj1} 

\begin{eqnarray}
0 &=&\left. \frac{d^2Q[\rho ({\bf r+}t{\bf R})]}{dt}\right| _{t=0}
\label{eqtra} \\
&=&\int {\bf R.\nabla }\rho ({\bf r})\frac{\delta ^2A[\rho ]}{\delta 
\rho (%
{\bf r})\delta \rho ({\bf r}^{\prime })}{\bf R.\nabla }^{\prime }\rho 
({\bf r%
}^{\prime })\ d^3rd^3r^{\prime }+\int \left[ \frac{\delta A[\rho 
]}{\delta
\rho ({\bf r})}\right] {\bf R.\nabla }\left[ {\bf R.\nabla }\rho 
({\bf r}%
)\right] \ d^3r.  \nonumber
\end{eqnarray}
Now, since $A[\rho ]$ is invariant under translational and ${\bf R}$ 
is
arbitrary, it follows that

\begin{equation}
\int \frac{d\rho ({\bf r})}{dx_i}\frac{\delta ^2A[\rho ]}{\delta \rho 
({\bf r%
})\delta \rho ({\bf r}^{\prime })}\frac{d\rho ({\bf r}^{\prime })}{%
dx_i^{^{\prime }}}\ d^3rd^3r^{\prime }=-\int \left[ 
\frac{d^2}{dx_i^2}\frac{%
\delta A[\rho ]}{\delta \rho ({\bf r})}\right] \rho ({\bf r})\ d^3r.
\label{eq10}
\end{equation}
The convex nature of $A[\rho ]$ combined with translational 
invariance
therefore imposes, from Eqs. (\ref{eq9}) and (\ref{eq10}), a 
constraint on
the first functional derivative of $A[\rho ]:$%
\begin{equation}
\int \left[ \frac{d^2}{dx_i^2}\frac{\delta A[\rho ]}{\delta \rho 
({\bf r})}%
\right] \rho ({\bf r})\ d^3r\leq 0  \label{eq11}
\end{equation}
Using the expression \cite{levygorling95} 
\begin{eqnarray}
\frac{\delta E_{xc}[\rho _\lambda ]}{\delta \rho ({\bf r})} &=&\int 
\frac{%
\delta E_{xc}[\rho _\lambda [\rho ]]}{\delta \rho _\lambda ([\rho 
];{\bf r}%
^{\prime })}\frac{\delta \rho _\lambda ([\rho ];{\bf r}^{\prime 
})}{\delta
\rho ({\bf r})}d^3r^{\prime }  \label{eq12} \\
&=&\int v_{xc}([\rho _\lambda [\rho ];{\bf r}^{\prime })\lambda 
^3\delta (%
{\bf r}-\lambda {\bf r}^{\prime }) \\
&=&v_{xc}([\rho _\lambda ;\frac{{\bf r}}\lambda ),
\end{eqnarray}
it follows from the definition of $A[\rho ]$ and Eq. (\ref{eq11}) 
that

\begin{equation}
\int \rho ({\bf r})\frac{d^2}{dx_i^2}\left[ v_H([\rho ],{\bf r}%
)+\lim_{\lambda \rightarrow 0}\frac 1\lambda v_{xc}([\rho _\lambda 
],\frac{%
{\bf r}}\lambda )\right] \ d^3r\leq 0,  \label{eq13}
\end{equation}
where $v_H(\left[ \rho \right] ,{\bf r})=\frac{\delta U[\rho 
]}{\delta \rho (%
{\bf r})}=\int \frac{\rho ({\bf r})}{\left| {\bf r-r}^{\prime 
}\right| }%
d^3r. $ Eq. (\ref{eq13}) imposes a constraint on $\frac 1\lambda
v_{xc}([\rho _\lambda ],\frac{{\bf r}}\lambda )$ in the low density 
limit.

Substituting $\nabla ^2$ for $\frac{d^2}{dx_i^2}$ Eq. (\ref{eq13}) is 
also
valid and the constraint can be expressed as$:$

\begin{eqnarray}
&&\int \rho ({\bf r})\nabla ^2\left[ v_H([\rho ],{\bf 
r})+\lim_{\lambda
\rightarrow 0}\frac 1\lambda v_{xc}([\rho _\lambda ],\frac{{\bf 
r}}\lambda
)\right] \ d^3r  \label{eq14} \\
&=&-4\pi \int \rho ({\bf r})^2\ d^3r+\lim_{\lambda \rightarrow 0}\int 
\rho (%
{\bf r})\nabla ^2\frac 1\lambda v_{xc}([\rho _\lambda ],\frac{{\bf 
r}}\lambda
)\ d^3r\leq 0  \nonumber
\end{eqnarray}
If we separate the correlation and exchange potentials, and take into
account that $v_x([\rho _\lambda ],\frac{{\bf r}}\lambda )\ =\lambda
v_x([\rho ],{\bf r})\ $\cite{huilevy91} Eqs. (\ref{eq13}) and 
(\ref{eq14})
may be written as 
\begin{equation}
\int \rho ({\bf r})\frac{d^2}{dx_i^2}\left[ v_H([\rho ],{\bf 
r})+v_x([\rho ],%
{\bf r})+\lim_{\lambda \rightarrow 0}\frac 1\lambda v_c([\rho 
_\lambda ],%
\frac{{\bf r}}\lambda )\right] \ d^3r\leq 0,  \label{eq13a}
\end{equation}
and 
\begin{equation}
\int \rho ({\bf r})\nabla ^2\left[ v_x([\rho ],{\bf r})+\lim_{\lambda
\rightarrow 0}\frac 1\lambda v_c([\rho _\lambda ],\frac{{\bf 
r}}\lambda )\
\right] d^3r\leq 4\pi \int \rho ({\bf r})^2\ d^3r,  \label{eq14a}
\end{equation}
respectively. In the special case of two electrons, $v_x([\rho ],{\bf 
r})=-%
\frac 12v_H([\rho ],{\bf r}).$ Hence, for this special case the 
constraint
on $v_c$ alone is 
\[
\int \rho ({\bf r})\nabla ^2\lim_{\lambda \rightarrow 0}\frac 
1\lambda
v_c([\rho _\lambda ],\frac{{\bf r}}\lambda )\ d^3r\leq 2\pi \int \rho 
({\bf r%
})^2\ d^3r 
\]

In the local density approximation 
\begin{equation}
A^{LDA}[\rho ]=-\left| c\right| \int \rho ({\bf r})^{\frac 43}\ 
d^3r+U[\rho
],  \label{eq15}
\end{equation}
Eq. (\ref{eq14}) requires that 
\begin{equation}
-4\pi \int \rho ({\bf r})^2\ d^3r-\frac 43\left| c\right| \int \rho 
({\bf r}%
)\nabla ^2\rho ({\bf r})^{\frac 13}\ d^3r\leq 0.  \label{eq16}
\end{equation}

This condition is easily violated. For example, with $\rho ({\bf 
r})=\chi
_m(1+\ r)^me^{-\ r},$ with $\chi _m$ a normalization constant, the 
value of $%
\left| c\right| $ for which Eq. (\ref{eq16}) is violated for can be 
made
arbitrarily small by increasing the power $m.$ Eq. (\ref{eq13}) thus 
imposes
a non-trivial constraint on the exchange-correlation potential.

\end{document}